# IZM-DSR: A New Zone-Disjoint Multi-path Routing Algorithm for Mobile Ad-Hoc Networks

Nastooh Taheri Javan
Amirkabir University of Technology
Tehran, Iran
nastooh@aut.ac.ir

Amir Dareshoorzade
Iran University of Science and Technology
Tehran, Iran
darehshoori@comp.iust.ac.ir

Saeed Soltanali
Amirkabir University of Technology
Tehran, Iran
ssoltanali@aut.ac.ir

Yasna Ghanbari Birgani
Azad University
Mahshahr, Iran
Yasna_ghanbari@yahoo.com

***Abstract*— Some of multi-path routing algorithms in MANETs use multiple paths simultaneously. These algorithms can attempt to find node-disjoint to achieve higher fault tolerance. By using node-disjoint paths, it is expected that the end-to-end delay in each case should be independent of each other. However, because of natural properties and medium access mechanisms in ad hoc networks, such as CSMA/CA, the end-to-end delay between any source and destination depends on the pattern of communication in the neighborhood region. In this case some of the intermediate nodes should be silent to reverence their neighbors and this matter increases the end-to-end delay. To avoid this problem, multi-path routing algorithms can use zone-disjoint paths instead of node-disjoint paths. In this paper we propose a new multi-path routing algorithm that selects zone-disjoint paths, using omni-directional antenna. We evaluate our algorithm in several different scenarios. The simulation results show that our approach is very effective in decreasing routing overhead and end-to-end delay.**

## I. INTRODUCTION

In ad hoc wireless networks, mobile nodes communicate without any fixed and preset infrastructure. Mobile nodes can roam arbitrarily without spatial or temporal constraints. To provide end-to-end communication throughout the network, each mobile node acts as an intermediate router forwarding messages received by other nodes [1]. The wireless radio link may be interrupted owning to one of the mobile nodes moving out from the original radio radius, running out of battery or being turn off by the user. The routing path between sender and the receiver can also be fractured.

Design of efficient routing protocols is the central challenge in such dynamic wireless networks. Routing protocols for MANETs can be broadly classified into reactive (on-demand) and proactive algorithms [2]. In reactive protocols, nodes build and maintain routes as they are needed but proactive routing algorithms usually constantly update routing table among nodes. In on-demand protocols, nodes only compute routes when they are needed. Therefore, on-demand protocols are more scalable to dynamic, large networks. When a node needs a route to another node, it initiates a route discovery process to find a route.

Numerous well-studied ad hoc wireless routing protocols, such as Dynamic Source Routing (DSR) [3] or Ad hoc On-Demand Distance Vector Routing (AODV)[4], rebroadcast the "Path Discovery Messages" and seek another routing path. Newly discovered paths may become un-useful even before the start of routing if network topology changes too frequently. Moreover, the network topology may change again before the last topology updates are propagated to all intermediate nodes.

Among the on-demand protocols, multi-path protocols have a relatively greater ability to reduce the route discovery frequency than single path protocols [5]. On-demand multi-path protocols discover multiple paths between the source and the destination in a single route discovery. So, a new route discovery is needed only when all these paths fail. In contrast, a single path protocol has to invoke new route discovery whenever the only path from the source to the destination fails. Thus, on-demand multi-path protocols cause fewer interruptions to the application data traffic when routes fail. They also have the potential to lower the routing overhead because of fewer route discovery operations.

Multi-path Routing can provide some benefits, such as load balancing, fault-tolerance, and higher aggregate bandwidth. Load balancing can be achieved by spreading the traffic along multiple routes; this can alleviate congestion and bottlenecks. From fault tolerance perspective, multi-path routing can provide route resilience. Since bandwidth may be limited in a wireless network, routing along a single path may not provide enough bandwidth for a connection, however, if multiple paths used simultaneously to route data, the aggregate of the paths may satisfy the bandwidth requirement of the application and a lower end-to-end delay may be achieved. Moreover, the frequency of route discovery is much lower if a node maintains multiple paths to destination.

After recognizing several paths between Source and destination in route discovery process in multi-path routing algorithms, data transferring can be started in co-process ways several routes. By using these mechanisms we can distribute load to several paths in order to balance the traffic, and increasing the bandwidth and as a result decreasing the delay.

Choosing the suitable paths for transferring the co-process data into the destination is the most important thing in this case. Choosing disjoint paths between source and destination is one of the ideas which is used in this process. This increases the fault tolerance noticeably.

As we know there are two problems in wireless networks, known as "hidden station" and "exposed station". For solving these problems, CSMA/CA [6] protocol has been suggested. In 802.11 standard, this protocol is used for access to the channel. Due to transferring RTS[2] and CTS[3] packets between nodes in this protocol, some of the nodes don't transfer the data and as a result the delay is increased.

As an example, consider figure 1 that shows an imaginary LAN with ten nodes [7]. In this figure radio range of every node is distinguished and the dotted line shows the relation between nodes. In other words, the dotted lines between two special nodes show that they are located in each other radio range.

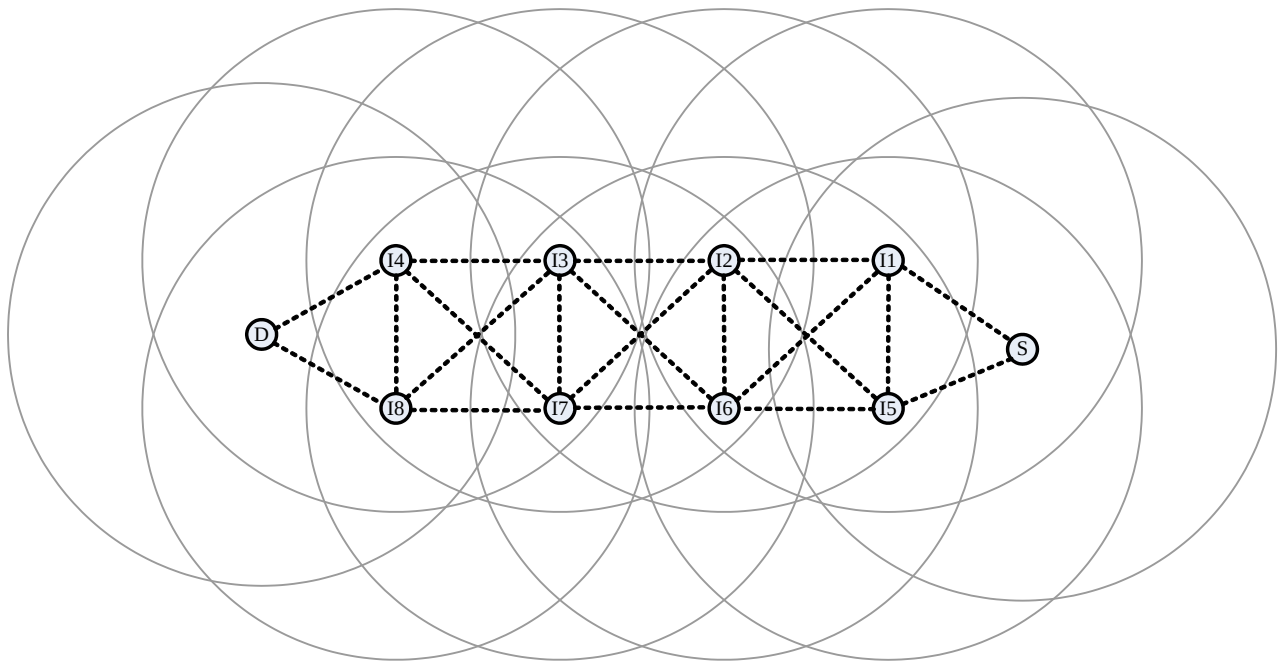

Figure 1. Node-Disjoin paths.

There are two node-disjoint paths, S-I1-I2-I3-I4-D and S-I5-I6-I7-I8-D, between D and S, which transferring the data in one path is not completely separated from the other path. In this case, the delay of every path is related to the other path traffic Because of transferring RTS[2] and CTS[3] packets between nodes of network in order to collision avoidance and solve hidden station and exposed station problems. As a result some of the station of a path in order to receive CTS from a node in opposite path should postpone their sending.

To solve this problem, we can use zone-disjoint paths instead of node-disjoint paths. Two routes with no pair of neighbor nodes are called zone-disjoint in terminology. In [8,9] the authors proposed a method for distinguishing the zone-disjoint paths, that uses directional antenna, but most of the present equipment is not equipped with directional antenna. In this paper a multi-path routing algorithm has suggested. In this approach, by using omni-directional antenna, the zone-disjoint paths are recognizable and can be used to send co-process data simultaneously.

A Multi-Path Routing Algorithm by the name of ZD-MPDSR has been offered in [7]. ZD-MPDSR discovers Zone-Disjoint paths between source and destination nodes, and source node uses these paths for simultaneous sending of date to destination. Although that algorithm gets some improvement of decreasing of end-to-end delay, but the overhead of its routing is so height. On the one hand, its route discovery process do with height delay but this delay is being compensated in data sending phase.

In this paper we improved the route discovery process in ZD-MPDSR which caused the delay of route discovery to be decreased.

The rest of this paper is organized as follows. The following section deals with the related works. Section III describes the proposed protocol mechanism in detail. Performance evaluation by simulation is presented in section IV and concluding remarks are made in section V.

## II. RELATED WORKS

Multi-path routing and its application have been well studied for wireless ad hoc networks.

The goal of SMR [10] is finding maximally disjoint multiple paths. SMR is an on-demand multi-path source routing that is similar to DSR [3]. To discovery the routing paths, the source, at first, broadcasts the RREQ to every neighbor. When the RREQ is delivered to a node, the intermediate node's ID is included into packet. Then node, receiving RREQ, re-broadcasts it to every outgoing path. In this algorithm, the destination sends a RREP for the first RREQ it receives, which represents the shortest delay path. The destination then waits to receive more RREQs. From the received RREQs, the path that is maximally disjoint from the shortest path is selected and destination sends a RREP fore the selected RREQ. In SMR the intermediate nodes do not reply to RREQs, this is to allow the destination to receive all the routes so that it can select the maximally disjoint paths.

AOMDV [12] is an extension to AODV [4] protocol for computing multiple loop-free and link-disjoint paths. In AOMDV through a modified route discovery process multiple link-disjoint paths are computed. The destination responds to only those unique neighbors from which it receives a route request. Each node in the network maintains a list of alternate next hops that are stored based on the hop count. If during routing, one of the links between two nodes break, then the immediate upstream node switches to the next node in its list of next hops. In this algorithm, the source node initiates a route request when all its alternate paths fail. The main drawback of this protocol is that the alternate paths that are computed during route discovery are not maintained during the course of data transfer.

In [8,9] multi-path routing with directional antenna is proposed. In this protocol directional antenna is used for finding zone-disjoint paths between a source and destination. Due to low transmission zone of directional antenna, it is easier to get two physically close paths that may not interfere with each other during communication.

Destination in both ZD-MPDSR[7] and ZD-AOMDV[10] tries to choose the zone-disjoint paths from received RREQs and send the RREPs to the source for these RREQs. In ZD-MPDSR, for recognizing zone-disjoint paths between source and destination, a new field is established in RREQ packet, which is called ActiveNeighborCount and it is initiated by zero. As a matter of fact this field shows the number of active neighbors for nodes on a path. Active neighbor is the node which received this RREQ and the source and destination may choose another path which has this node on it, and in this case sending the data from selected paths, is related to each other. In order to set the ZD-MPDSR up, the entire nodes should keep a table which is called RREQ_Seen and this table records the characteristics of received RREQs by every node.

In ZD_MPDSR intermediate node shouldn't send RREP to any source and in fact should let the destination receive all RREQs and choose the best paths and send RREPs to the source. In other words in ZD-MPDSR, the intermediate nodes don't need Route Cache.

In this algorithm, the source node initiates and floods a RREQ packet in order to recognize a route to destination. As mentioned before, initiate value of ActiveNeighborCount in this packet is zero. In this case every intermediate nodes which received a RREQ, records it's characteristics in RREQ_Seen table, but before sending this packet, asks its neighbors "Have you seen this RREQ with this characteristics before?" and sends a packet which is called RREQ_Query to its neighbors and waits for the reply for specified time distinguished by a timer. In this case the neighbors have to reply the answer by searching in their RREQ_Seen table. When the time is up, this node increases the value of ActiveNeighborCount in RREQ packet amount of number of neighbors that send positive answer, and then flood RREQ packet to neighbors.

In this case when the destination received different RREQs, starts to choose separated paths and then between chosen paths considers values of ActiveNeighborCount and chooses the paths which have less values of ActiveNeighborCount. In fact the destination by choosing the paths which have less values of ActiveNeighborCount, tries to select the zone-disjoint paths. Then destination sends the RREP packets to source from chosen paths. As soon as the source receives the first RREP starts to transfer the data by this route and after receiving the next RREP, divides the load into the present routes based on criteria which are about load balancing. These criteria are exchangeable in the way of distributing the load in the routes.

## III. Proposed Algorithm

In fact, our proposed algorithm is an improvement of ZD-MPDSR. ZD-MPDSR that is based on DSR uses several Zone-Disjoint paths between source and destination to send data over multiple paths simultaneously.

In this section, first we consider the same properties of proposed algorithm and ZD_MPDSR, then we propose the details of our new algorithm.

Like ZD-MPDSR, intermediate nodes do not need any Route Cache Table. The RREP packet in both ZD_MPDSR and IZM-DSR need the ActiveNeighborCount field to find zone-disjoint paths. Like ZD_MPDSR, in IZM-DSR intermediate nodes need RREQ_Seen table with a few changes.

### *A. Difference Between IZM-DSR and ZD-MPDSR*

In this section we consider some using changes in IZM-DSR versus ZD-MPDSR.

In our new algorithm, against ZD-MPDSR, the RREP packets do not need the ActiveNeighborCount field. IZM-ZDS do not need any RREQ_Query and RREQ_Query_Reply packet for finding the active neighbors. Against ZD-MPDSR, that source selects paths, in IZM-DSR, destination selects paths for sending data simultaneously. The RREQ_Seen tables in intermediate nodes have an extra field which name is Counter to account the number of received RREQs.

### *B. IZM-DSR algorithm*

In on-demand algorithms, when source has data to send, but no route to the destination is known, source sends a route request packet to all its neighbors.

in IZM-DSR, each intermediate nodes that receive RREQ, insert this packet in RREQ_Seen Table and increase the Counter field in RREQ_Seend Table.

When destination receive RREQs, it creates a RREP with the Activeneighborcount value sets to zero, and send it to the source along the reverse path that included in RREQ.

Each intermediate nodes that receive a RREP packet use the count field in RREQ_Seen Table to update the the activeneighborcount field.

When a RREP receives to the source, the source wait for a certain time to receive all other RREPs. After that source can select some paths with less activeneighborcount field from received RREP and can send data over selected routes.

To understand the operation of nodes in this Algorithm, pseudo code of source node, intermediate nodes and destination node are shown in figure 2,3 and 4 respectively.

1. When no route information to the destination is known send RREQ.
2. Wait for return RREP packets from destination.
3. If return first RREP from destination with adjust a timer, wait specific duration for receiving rest of RREP packets.
4. After expired time of timer, ascending sort the received RREP packets with ActiveNeighborCount field exist on it.
5. According the consideration contract in Load Balancing field, to needed numbers, select path from first of sorting list of RREP packets.
6. Begin sending data to destination as simultaneous, from selected paths in step 5.

Figure 2. Pseudo code for the source node in IZM-DSR.

1. After receipt RREQ packets, according its Routing Algorithm, instead of each RREQ packet, send a RREP packet with primary amount of ActiveNeighborCount equal Zero to source.

Figure 3. Pseudo code for the intermediates node in IZM-DSR.

1. If receive a RREQ packet, and this packet is new and acceptable packet, insert characteristics of this RREQ packet in RREQ_Seen table and equal Zero Counter field of similar of it.
2. If receive a repetitive RREQ packet, increase one unit, Counter field regarding to this RREQ in RREQ_Seen table.
3. According politic of Route Discovery, if RREQ packet is able to re-broadcast, broadcast it for all.

Figure 4. Pseudo code for the destination node in IZM-DSR.

Figure 5 shows an example of our new proposed algorithm. If node S wants to send data to D and no route information is known such as DSR it broadcasts a RREQ. Each intermediate nodes that receive this RREQ increase the counter filed in its RREQ-Seen Table and broadcast it until it reaches to the Destination.(figure 6)

Imagine that in figure5, node S wants synchronous transferring data with two paths for node D. So with broadcast sending of RREQ packet, begin route discovery process. According the principles of DSR Algorithm, RREQ packets go through their way to destination and in this process, middle nodes update Counter field in their RREQ_Seen table with receiving every RREQ.

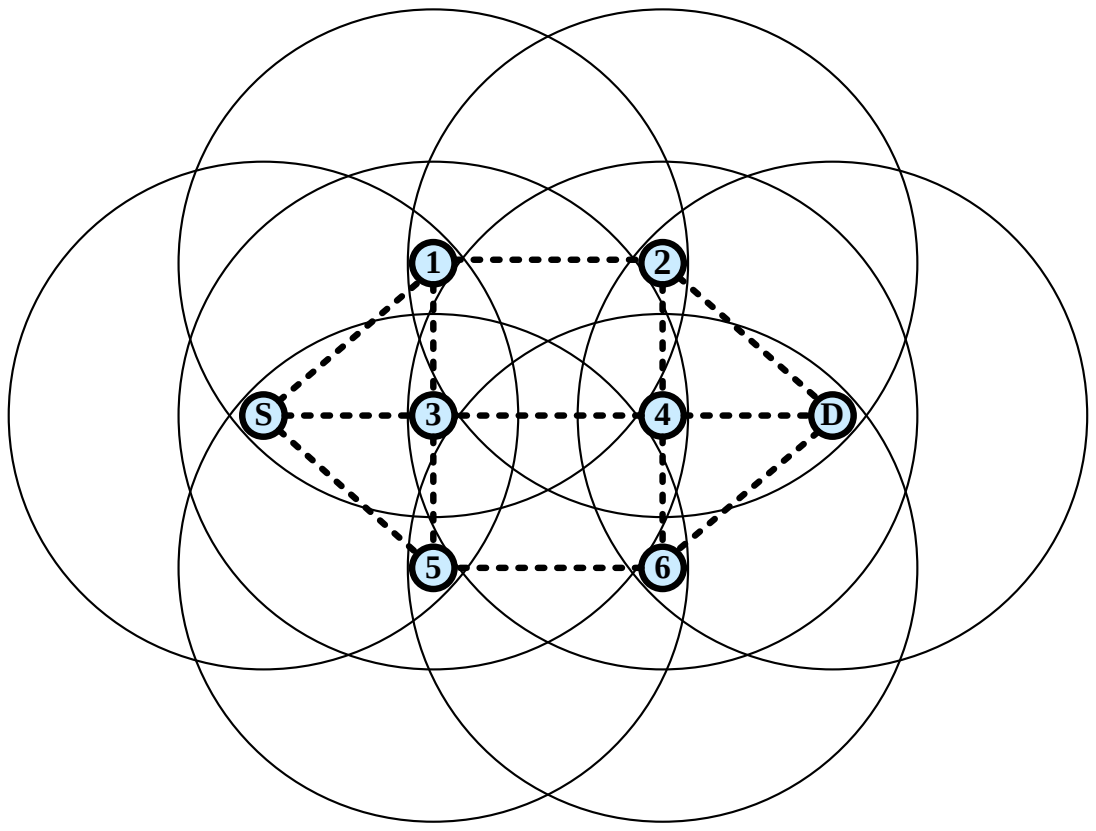

Figure 5. An Example for proposed algorithm..

Figure6 shows the state of this network when all of RREQs receive to destination. As shown in figure 6 since node 1 receives two RREQ from node 2 and 3 the counter field in its RREQ_Seen table sets to two.

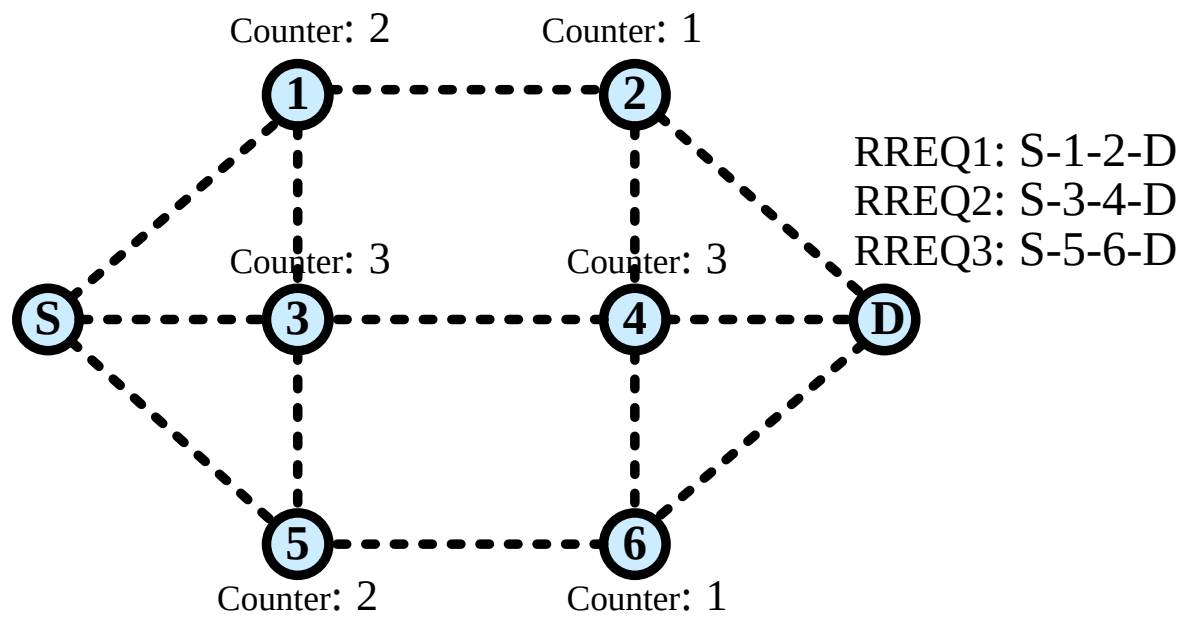

Figure 6. Destination receive RREQs in IZM-DSR.

When destination receives a RREQ it creates a RREP and sets the activeneighborcount(A_N_C) field to zero. In reverse path each intermediate nodes that receive the RREP add the value of Count in RREQ_Seen Table to the activeneighborcount field in RREP.(figure 7)

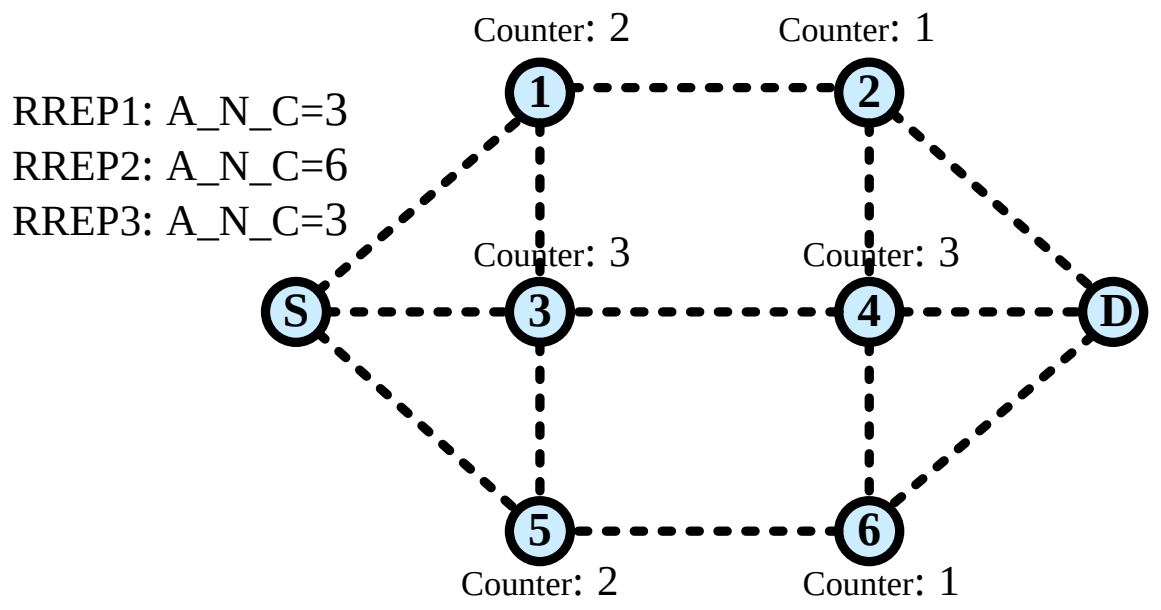

Figure 7. Source node receive RREPs in IZM-DSR.

As shown in figure 7 three RREPs receive to the source and since the first and third path have less neighboractivecount in compare of second path thus source select S-1-2-D and S-5-6-D as the discovered routes for sending data simultaneously along these paths.

## IV. Performance Evaluation

In order to demonstrate the effectiveness of our algorithm, we evaluate our proposed protocol and compare its performance to SMR[10] and ZD-MPDSR[7].

### A. *Simulation Environment*

Our simulation runs on the GloMoSim simulation platform [12]. The GloMoSim library is a detailed simulation environment for wireless network systems.

Our simulation environment consists of N mobile nodes in a rectangular region of size 1000 meters by 1000 meters. The nodes are randomly placed in the region and each of them has a radio propagation range of 250 meters.

The radio model to transmit and receive packets is RADIO-ACCNOISE which is the standard radio model used. The IEEE 802.11 was used as the medium access control protocol.

The random waypoint model was adopted as the mobility model. In the random waypoint model, a node randomly selects a destination from the physical terrain. It moves in the direction of the destination in a speed uniformly chosen between a minimum and maximum speed specified. After it reaches its destination, the node stays there for a time period specified as the pause time. In our simulation, minimum speed was set constant to zero.

The simulated traffic is Constant Bit Rate (CBR) and all data packets are 512 Bytes. Each simulation is run for 300 seconds.

### B. *4.2. Performance metrics*

The following metrics are used to evaluate the performance: (i) Packet delivery Ratio – the ratio between the number of data packets received and those originated by the sources. (ii) Routing overhead – the total control packet transmitted by each node. (iii) Average end-to-end delay – the time from when the source generates the data packet to when the destination receives it. This includes: route acquisition latency, processing delays at various layers of each node, queuing at the interface queue, retransmission delays at the MAC, propagation and transfer times.

### *C. Simulation Scenarios*

We ran experiments with two different base settings. In the first setting, 100 nodes are randomly placed inside an area of 1000 x 1000 m2.

For the second setting, the number of nodes and the size of the simulation area are varied, while keeping the average node density constant.

### *D. Simulation Results*

In the first scenario, to evaluate capability of the protocols for different node mobility, we change node mobility by varying the maximum speed. The number of nodes and pause time was fixed at 100 nodes and 1 second, respectively.

Figures 8 and 9 show the evolution of the delivery ratio and average delay for increasing node speed (from 10 to 50 m/s) in a random waypoint model.

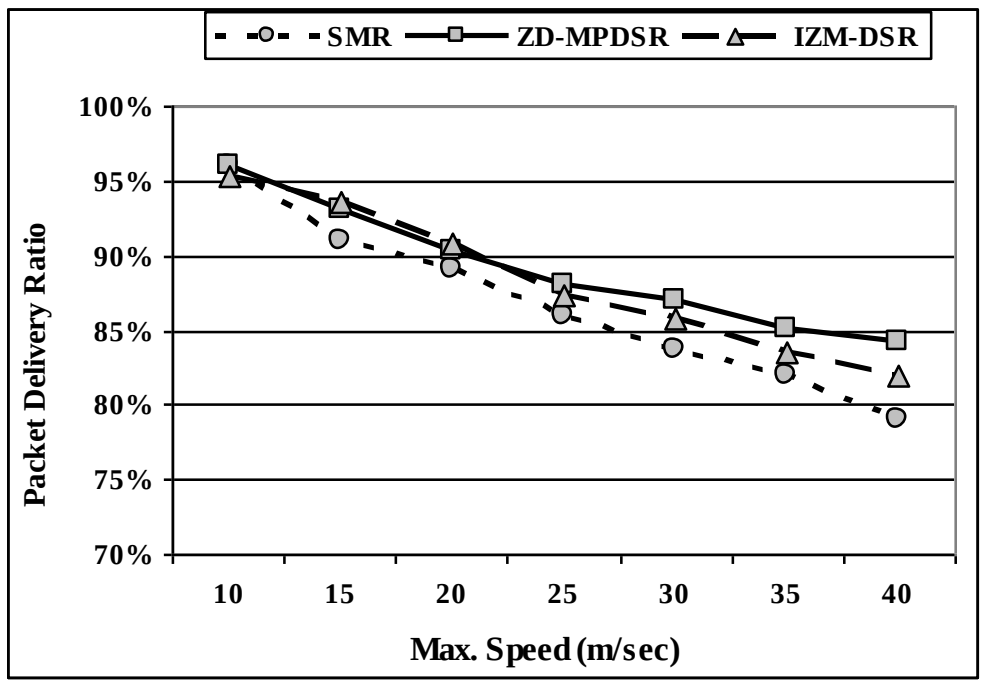

Figure 8. The packet delivery ratio with varying speed.

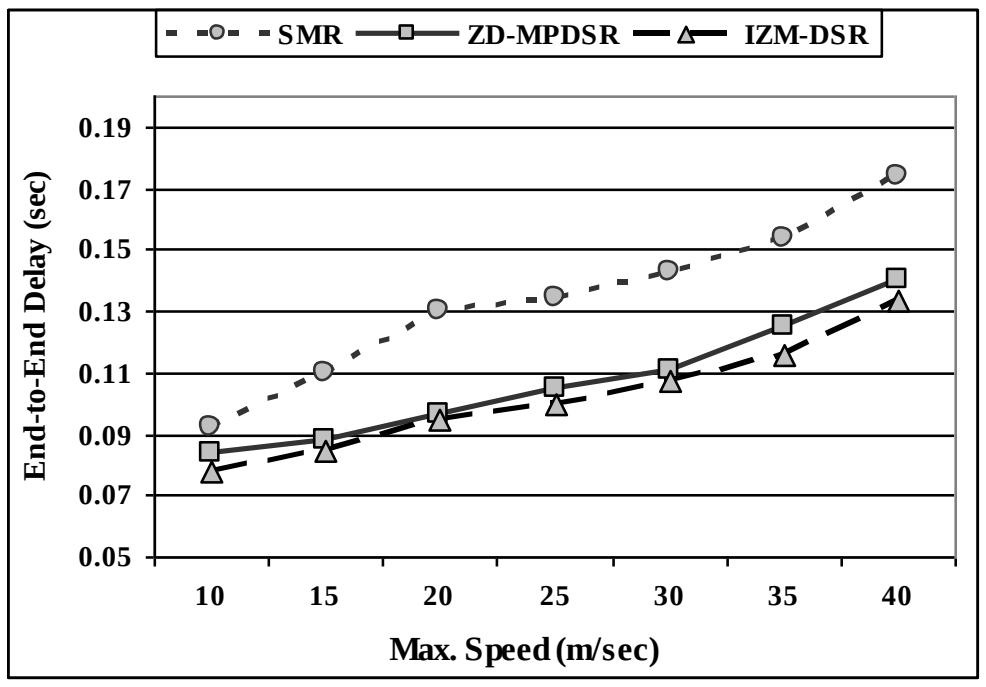

Figure 9. The average end-to-end delay with varying speed.

Since in both ZD-MPDSR and IZM-DSR zone disjoint paths are used for sending data the end-to-end delay is less than SMR. In this scenario the IZM-DSR exhibits the lower end-to-end delay than the ZD-MPDSR (Fig. 9), and also has greater packet delivery ratio than ZD-MPDSR (Fig. 8).

Figure 10 shows the overhead in the considered test scenarios with variable speed for Random way point mobility model. Overhead is expressed as the number of control packets forwarded. Since the ZD_MPDSR uses the RREQ_Query and RREQ_Query_Reply in route discovery, its overhead is more than DSR and SMR.

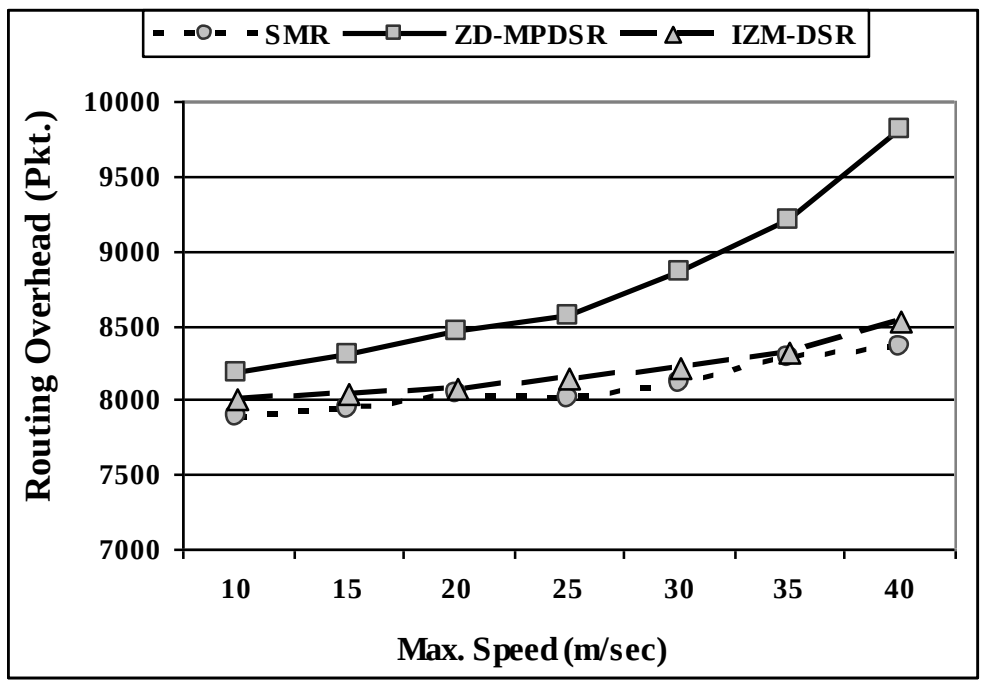

Figure 10. The routing overhead with varying speed.

In the Second set of experiments, of which the results are visualized in figures 11 to 13, we study the three algorithms in large network simulations. In this scenario, the number of nodes and the size of the simulation area are varied, while keeping the average node density constant.

In this case, IZM-DSR has lower end-to-end delay than ZD-MPDSR and SMR (Fig. 11), and also exhibits greater packet delivery ratio than ZD-MPDSR and SMR (Fig. 12).

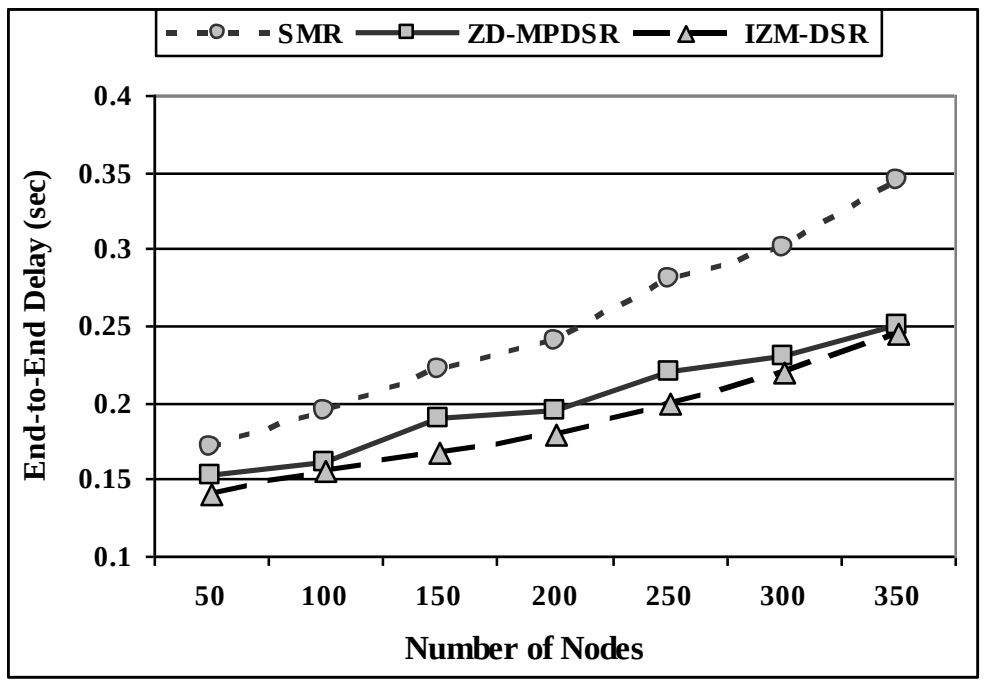

Figure 11. The average end-to-end with varying network size.

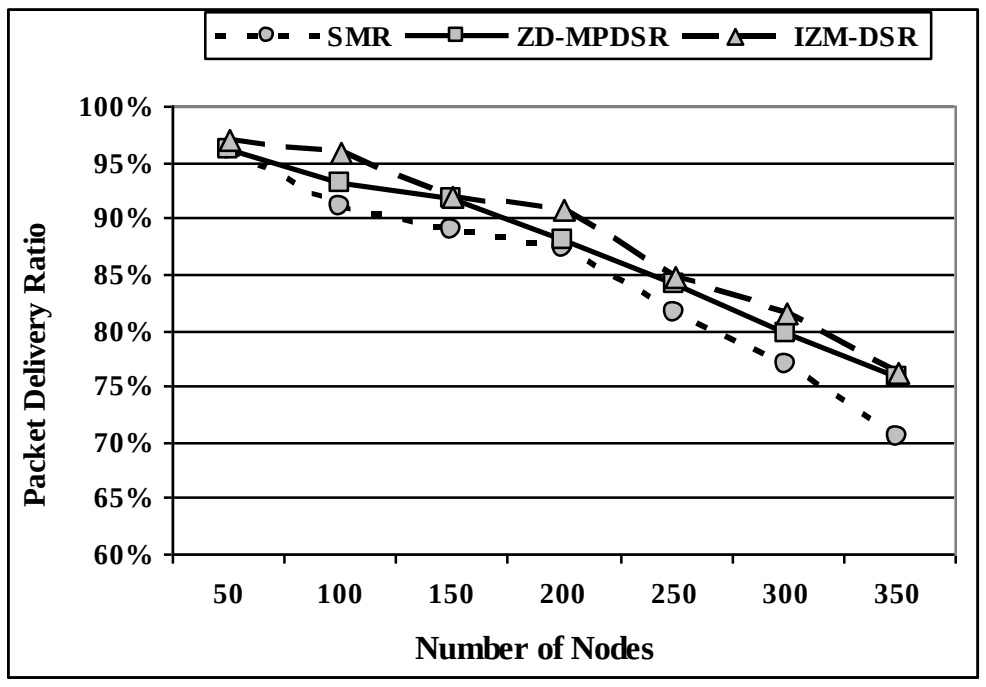

Figure 12. The packet delivery ratio with varying network size.

The experimental results represent that if the size of network increase, the amount of receiving data packets by destination nodes will be decrease.

In this scenario, the ZD-MPDSR has greater control overhead than IZM-DSR (Fig. 13).

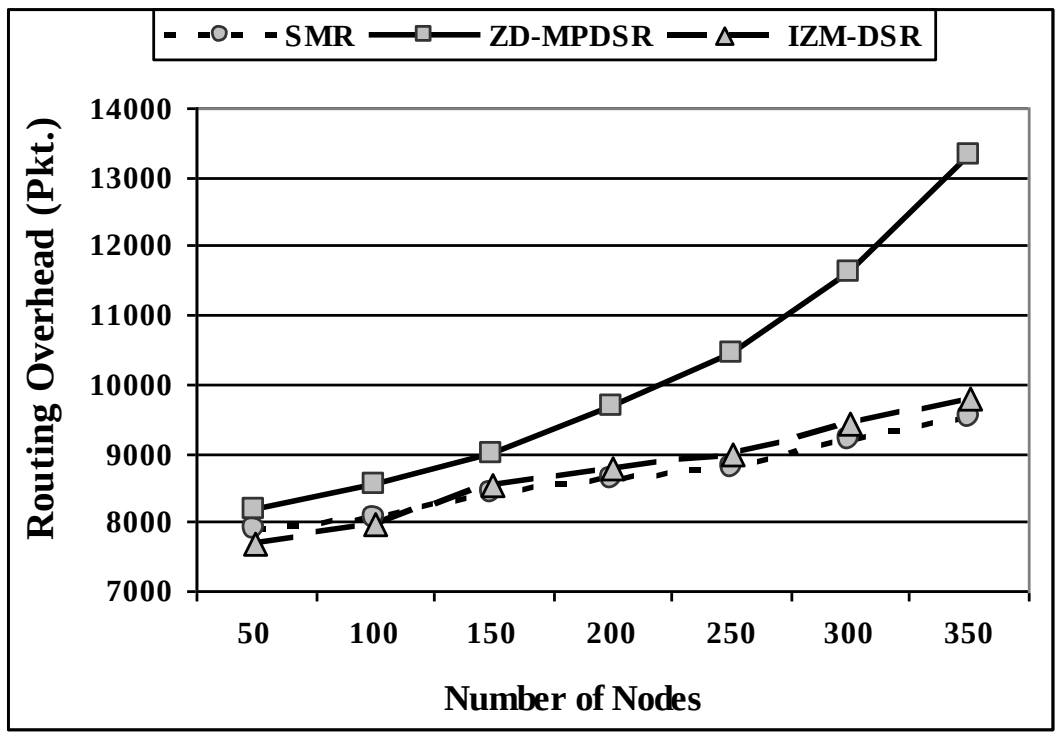

Figure 13. The routing overhead with varying network size.

## V. CONCLUSION

Some of the multi-path routing algorithms in MANET in order to decline the end-to-end delay divide the data in the source and send different sections to the destination by several routes simultaneously. In this way using the node-disjoint routes is the best option, but even sending data by the node-disjoint routes are not separated from each other and due to nature of MANET and CSMA/CA protocol, sending data by a route affect on the other route. In this paper a new multi-path routing algorithm is presented, in which by using common and omni-directional antenna can recognize the zone-disjoint routes between two nodes and use these routes to send the co-process data. The simulation results show that proposed algorithm is very effective in decreasing routing overhead and also decreasing the end-to-end delay in MANETs.